\documentclass[aps,pra,superscriptaddress,twocolumn,amsmath,amssymb]{revtex4-2}
\pdfpageattr {/Group << /S /Transparency /I true /CS /DeviceRGB>>}

\usepackage{graphicx} 
\usepackage{color}
\usepackage[hidelinks]{hyperref}
\usepackage{orcidlink}
\usepackage{amsmath,dsfont}
\usepackage{amsthm}
\usepackage{physics}
\usepackage{amsfonts}
\usepackage{times,txfonts}
\usepackage[normalem]{ulem}
\hypersetup{colorlinks=true, linkcolor=black, citecolor=black, urlcolor=blue}

\begin{document}
\title{Input-Output Analysis of Quantum Dot SUPER Excitation}

\author{Johannes Kerber\,\orcidlink{0009-0002-1957-8008}}
\thanks{These authors contributed equally to this work.}
\affiliation{Institut für Theoretische Physik, Universität Innsbruck,
Technikerstraße 21a, A-6020 Innsbruck, Austria}
\email{Johannes.Kerber@uibk.ac.at}

\author{Christoph Hotter\,\orcidlink{0009-0003-3854-0264}}
\thanks{These authors contributed equally to this work.}
\affiliation{Niels Bohr Institute, University of Copenhagen,
Jagtvej 155A, Copenhagen DK-2200, Denmark}
\affiliation{Department of Physics, MIT-Harvard Center for Ultracold Atoms
and Research Laboratory of Electronics, Massachusetts Institute of Technology,
Cambridge, Massachusetts 02139, USA}
\email{hottch@mit.edu}

\author{Helmut Ritsch\,\orcidlink{0000-0001-7013-5208}}
\affiliation{Institut für Theoretische Physik, Universität Innsbruck,
Technikerstraße 21a, A-6020 Innsbruck, Austria}

\author{Klaus Mølmer\,\orcidlink{0000-0002-2372-869X}}
\affiliation{Niels Bohr Institute, University of Copenhagen,
Jagtvej 155A, Copenhagen DK-2200, Denmark}

\date{\today}

\begin{abstract}
 The Swing-UP of the quantum EmitteR population (SUPER)  two-color pulsed excitation scheme allows for robust and close to 100\% excitation of a two-level quantum emitter using only red detuned light. We analyze the underlying counterintuitive dynamics using a full quantum input-output description of the in- and outgoing light pulses for the case of two coherent free-space input pulses at realistic photon numbers. At the microscopic level, the SUPER mechanism exhibits its nonlinear three-photon Raman-type character, leading to a net photon-number change of $-2$ in one mode and $+1$ in the other, as was first guessed from cavity-enhanced model descriptions at low photon numbers. We confirm that in free space, a sufficiently high pulse photon number, much larger than one, is required to achieve high-fidelity inversion. To treat the large coherent-state amplitudes (photon numbers) relevant for SUPER, we extend the quantum input-output formalism to include a cumulant expansion approach. With an interaction-picture formulation, the few exchanged photons that govern the nontrivial dynamics enable direct full-quantum calculations in truncated Hilbert spaces, including treatments in a displacement frame and for initial Fock-state pulses.
\end{abstract}
\maketitle


\section{Introduction}
High-purity deterministic single-photon generation requires both an emitter platform of high quantum efficiency and a deterministic excitation protocol to prepare the excited state with high fidelity. For semiconductor quantum dots, the recently introduced \textbf{S}wing-\textbf{UP} of the quantum \textbf{E}mitte\textbf{R} population scheme (SUPER)~\cite{bracht_swing-up_2021}, which relies on two carefully designed time-overlapping, far off-resonant red detuned Gaussian pulses, was shown to generate such deterministic high fidelity population inversion with negligible background bulk excitation and emitter re-excitation. The large red detunings employed in the SUPER scheme place the driving laser several meV below the quantum-dot excitonic resonance, minimizing coupling to higher-energy states while enabling fully coherent excitation and efficient spectral separation of the emitted photons from the driving laser \cite{bracht_swing-up_2021, karli_super_2022}. The somewhat counterintuitive underlying mechanism has already been extensively analyzed in semiclassical models exhibiting a highly complex time evolution of the Bloch vector ending up in the inverted state. The method proves robust even in various generalized settings, including multi-level emitters, dissipative environments, and collectively coupled dot pairs~\cite{bracht_swing-up_2021,karli_super_2022,bracht_dressed_2023,heinisch_swing_2024,kerber_selective_2026,nunner_collective_2026, crowder2026quantumdot, piccinini2025excitonbiexcitonpreparationcoherent, hagen2026}, but its basic underlying mechanisms remained somewhat obscure. 

As a single photon from either of the two red-detuned pulses cannot provide enough excitation energy for the dot, one expects genuine multi-photon dynamics to emerge. Different simplified two-mode Jaynes-Cummings models with strong coupling of the cavity photons to the emitter have given insights into how this off-resonant two-color excitation occurs~\cite{richter_few-photon_2025, Vannucci2024}. In essence, the process is qualitatively tied to a generalized 3-photon Raman process, where two photons from the higher frequency field of the first pulse are absorbed, and one lower frequency photon is added to the second pulse. Important contributions in the dynamics also originate from time-dependent Stark shifts induced by the two pulses, which are not so easy to capture in the few-photon limit. Hence, a detailed quantum description of SUPER for large amplitude coherent traveling pulses, which is essential to analyze the quantum states of the generated fields, is still missing, as an effective time-dependent coupling to discrete cavity modes falls short of reproducing the full dynamics of propagating quantum pulses~\cite{christiansen_interactions_2023}.

In the alternative approach we are following here, the underlying free space pulse dynamics are explicitly described within the recently developed input-output temporal mode description for quantum light pulses~\cite{kiilerich_input-output_2019,kiilerich_quantum_2020}. This can be nicely formulated in combination with the SLH formalism for quantum networks~\cite{combes_slh_nodate}. Essentially, it generalizes standard input-output theory~\cite{gardiner_input_and_output} by representing the temporal modes of incident and outgoing pulses through auxiliary effective modes with suitably engineered time-dependent couplings. This approach provides direct access not only to the emitter dynamics, but also to the quantum state and temporal-mode structure of the scattered radiation~\cite{baragiola2012nphoton}.

In this work, we re-analyze the SUPER scheme within this pulse-based input-output framework and use it to identify its central quantum features. We show that the traveling-pulse realization of SUPER clearly exhibits the underlying multi-photon character. In contrast to a cavity system, described by a two-mode Jaynes-Cummings model~\cite{richter_few-photon_2025, Vannucci2024}, a free space geometry demands much higher photon numbers to achieve the required field amplitudes for high-fidelity inversion. 

We will also use this example as a testbed to introduce several complementary models, including a direct cascaded description, an interaction-picture formulation, and full-quantum treatments in truncated Hilbert spaces or the use of a displacement frame for large field amplitudes. For the large photon numbers of the coherent pulses relevant here, the cumulant expansion provides an efficient description of the system and mode dynamics, while the interaction-picture formulation clarifies that only a few exchanged photons govern the nontrivial excitation process. 

The manuscript is organized as follows. In Sec.~\ref{sec:model}, we introduce the classical reference model and the pulse-based input-output and SLH description of the SUPER scheme. In Sec.~\ref{sec:cascade}, we analyze the cascaded two-pulse problem for coherent Gaussian input pulses, including the interaction-picture formulation and the minimum-photon-number requirement. In Sec.~\ref{sec:quantum}, we present full quantum descriptions in a displacement frame as well as for incident Fock-state pulses. Additional technical details, including multimode couplings, concatenated and single-pulse representations, and the interaction-picture construction, are collected in the Appendices.


\section{Model} \label{sec:model}

\subsection{Classical Reference Model}
Let us first briefly recall the semiclassical description of the SUPER protocol~\cite{bracht_swing-up_2021} for a driven single two-level system (TLS) with transition frequency $\omega_0$ and spontaneous emission rate $\gamma$. In a rotating frame at the energy of the TLS, the Hamiltonian reads ($\hbar \equiv 1$)

\begin{align}
    \hat{H}_{\text{cl}}(t) = - \frac{1}{2}(\Omega_{\text{S}}(t)\hat{\sigma}^+ + \text{h.c.}),\label{eq.Ham_SUPER_2}
\end{align}
where $\hat{\sigma}^{\pm}$ are the TLS ladder operators. The coherent drive by two detuned Gaussian pulses generates a fixed time-dependent Rabi frequency of the form:

\begin{align}
    \Omega_{\text{S}}(t) = \Omega_1(t)e^{-i\Delta_1 t} + \Omega_2(t)e^{-i\Delta_2 t},\label{eq.SUPER}
\end{align}
with envelopes $\Omega_i(t) = \frac{A_i}{\sqrt{2\pi\sigma_i^2}} e^{-(t-\tau_i)^2/(2\sigma_i^2)}$ and fixed detunings $\Delta_i = \omega_i - \omega_0$. Here $A_i$ is the pulse area, $\sigma_i$ the pulse width, and $\tau_i$ the time offset. The full open systems dynamics is then governed by the Lindblad master equation

\begin{align}
    \partial_t\hat{\rho} = -i[\hat{H},\hat{\rho}] + \sum_{i=1}^{n}\mathcal{D}[\hat{L}_i]\hat{\rho},
    \label{eq.master}
\end{align}
with $\mathcal{D}[\hat{L}_i]\hat{\rho} = \hat{L}_i\hat{\rho}\hat{L}_i^{\dagger} - \frac{1}{2}(\hat{L}_i^{\dagger}\hat{L}_i\hat{\rho} + \hat{\rho}\hat{L}^{\dagger}_i\hat{L}_i)$. Here the only jump operator is $\hat{L}_{\text{cl}}=\sqrt{\gamma}\hat{\sigma}^-$. This equation can be readily numerically solved as a reference case for the refined input-output model studied below, which includes the dynamical evolution of the pulses introduced by the presence of the quantum emitter. 

\subsection{Input-Output Theory and SLH Formalism}\label{suse:IOTSLH}
To describe the excitation laser light as incoming pulses in a quantum description, we use the input-output theory with quantum pulses~\cite{kiilerich_input-output_2019,kiilerich_quantum_2020} embedded in the SLH framework for cascaded quantum networks~\cite{combes_slh_nodate}. In this approach, a propagating pulse mode is represented by launching a well-defined quantum state from a virtual cavity using a suitably engineered time-dependent coupling. The quantum properties of the light pulse are determined by the initial state in a virtual cavity: a cavity initially prepared in a coherent state $\ket{\alpha}$ emits a pulse in the chosen temporal mode with coherent amplitude $\alpha$ and mean photon number $|\alpha|^2$. A single photon state in the cavity creates a traveling single photon wave packet with a designable waveform. This mapping turns the usual scattering problem into a Markovian open-system description that simultaneously tracks the localized emitter and the selected pulse modes.

For a normalized temporal mode $u(t)$, the corresponding bosonic creation operator is~\cite{kiilerich_input-output_2019}
\begin{align}
    \hat{b}^{\dagger}_u = \int dt\,u(t)\hat{b}^{\dagger}(t),
\end{align}
with $[\hat{b}(t),\hat{b}^{\dagger}(t')] = \delta(t-t')$ and $\int dt\,|u(t)|^2=1$. For a single incident mode $u(t)$ and a single selected output mode $v(t)$, the virtual input and output cavities must have the time-dependent couplings~\cite{gough_generating_nonclassical_2015,christiansen_interactions_2023,kiilerich_input-output_2019,kiilerich_quantum_2020,lund_perfect_2023}
\begin{align}
    g_u(t) = \frac{u^*(t)}{\sqrt{1 - \int_0^t dt'|u(t')|^2}},
    \qquad
    g_v(t) = -\frac{v^*(t)}{\sqrt{\int_{0}^{t}dt'|v(t')|^2}}.
\end{align}
In the case of several cascaded input or output modes, the effective couplings must be modified to account for the pulse distortions induced by the subsequently following virtual cavities in the network~\cite{kiilerich_input-output_2019,kiilerich_quantum_2020}. The corresponding expressions are collected in Appendix~\ref{app:couplings}.

Within the SLH formalism, each network component is described by an operator triplet
\begin{align}
    G=(\hat{S},\hat{L},\hat{H}),
\end{align}
where $\hat{S}$ is the scattering matrix, $\hat{L}$ the coupling operator, and $\hat{H}$ the internal Hamiltonian. The triplet for a single virtual cavity is 
\begin{align}
    G_i=(\mathds{1},g_i(t)\hat{a}_i,0), 
\end{align}
while a single two-level emitter is described by
\begin{align}
    G_{\text{TLS}}=(\mathds{1},\sqrt{\gamma}\hat{\sigma}^-,-\Delta\hat{\sigma}^+\hat{\sigma}^-).
\end{align}
Throughout this manuscript, we assume zero reference detuning ($\Delta = 0$). The frequency differences between the photon pulses and the TLS are taken into account by the temporal modes. If two components $G_1=(\hat{S}_1,\hat{L}_1,\hat{H}_1)$ and $G_2=(\hat{S}_2,\hat{L}_2,\hat{H}_2)$ are connected in a cascade such that the output of $G_1$ drives $G_2$, the combined system is given by the SLH cascade product~\cite{Gough2009,gough_generating_nonclassical_2015,combes_slh_nodate}
\begin{align}
    G_2\lhd G_1 = \left(\hat{S}_2\hat{S}_1,\hat{L}_2+\hat{S}_2\hat{L}_1,\hat{H}_1+\hat{H}_2+\frac{1}{2i}\left(\hat{L}_2^{\dagger}\hat{S}_2\hat{L}_1-\hat{L}_1^{\dagger}\hat{S}_2^{\dagger}\hat{L}_2\right)\right).
\end{align}
This ensures that using this total triplet $G=(\hat{S},\hat{L},\hat{H})$, the density operator evolves according to the Lindblad master equation in Eq.~\eqref{eq.master}.

\subsection{Two-Input and Two-Output Cascade}

\begin{figure}
    \centering
    \includegraphics[clip, trim=0.5cm 20.0cm 1.5cm 2.0cm, width=.99\linewidth]{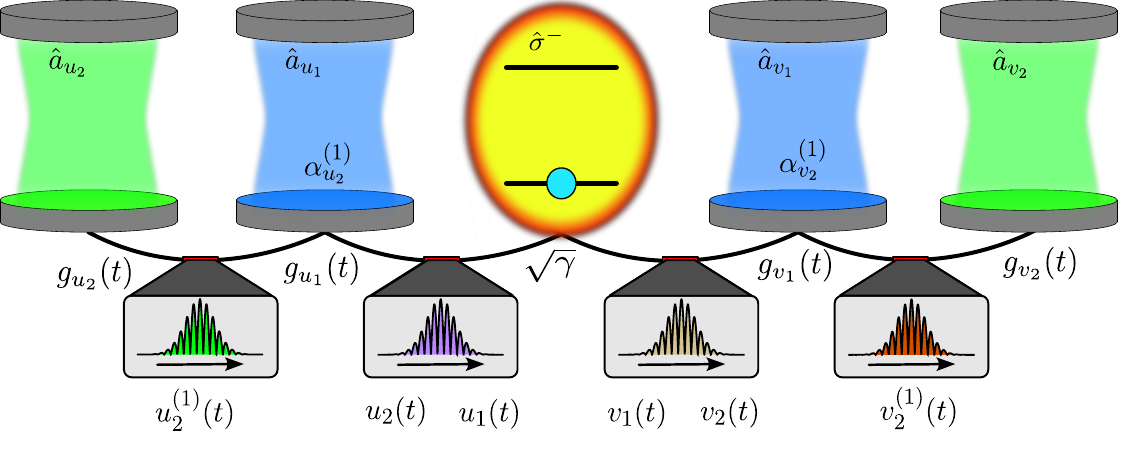}
    \caption{Cascade setup corresponding to the SLH triplet: $G^{\text{cas}}$ (from left to right). The virtual input cavities ($\hat{a}_{u_i}$, $g_{u_i}(t))$ emit the pulses $u_1(t)$ and $u_2^{(1)}(t)$, respectively. The distortion amplitude $\alpha^{(1)}_{u_2}$ modifies $u_2^{(1)}(t)$ such that $u_2(t)$ together with $u_1(t)$ interact with the TLS, simultaneously. Similarly for the virtual output cavities ($\hat{a}_{v_i}$, $g_{v_i}(t)$), where the outgoing pulse $v_2(t)$ is modified to $v_2^{(1)}(t)$ by the amplitude $\alpha_{v_2}^{(1)}$.}\label{fig.schem_1}
\end{figure}

The specific example of SUPER excitation that we study below considers two incident quantum pulses represented by two input modes and two correspondingly selected output modes arranged in a single cascade, see Fig.~\ref{fig.schem_1}. We choose the selected output modes to be coupled to the emitter using the same temporal shapes as the inputs, $v_i(t)=u_i(t)$. The network, therefore, contains four virtual cavities and the TLS. 
Starting from the RHS and applying the cascade rule recursively in the order shown in Fig.~\ref{fig.schem_1} yields
\begin{align}
    G^{\text{cas}}=G_{v_2}\lhd G_{v_1}\lhd G_{\text{TLS}}\lhd G_{u_1}\lhd G_{u_2},
    \label{eq.G_cas1}
\end{align}
leading to 
\begin{align}
    \hat{L}^{\text{cas}}(t) &= \sqrt{\gamma}\hat{\sigma}^- + \boldsymbol{g}^{\dagger}(t)\boldsymbol{a}, \label{eq.Lind}\\
    \hat{H}^{\text{cas}}(t) &= -\Delta\hat{\sigma}^{+}\hat{\sigma}^{-}
    + \frac{i}{2}\boldsymbol{a}^{\dagger}\boldsymbol{A}(t)\boldsymbol{a}
    - \frac{\sqrt{\gamma}}{2i}\left(\hat{\sigma}^{+}\tilde{\boldsymbol{g}}^{\dagger}(t)\boldsymbol{a} - \text{h.c.}\right).
    \label{eq.Ham}
\end{align}
Here, we define 
\begin{align}
    \boldsymbol{a} &= (\hat{a}_{u_2},\hat{a}_{u_1},\hat{a}_{v_1},\hat{a}_{v_2})^{T}, \\
    \boldsymbol{g} &= (g_{u_2}^*,g_{u_1}^*,g_{v_1}^*,g_{v_2}^*)^T, \\
    \tilde{\boldsymbol{g}} &= (g_{u_2}^*,g_{u_1}^*,-g_{v_1}^*,-g_{v_2}^*)^T
\end{align}
and
\begin{align}
    \boldsymbol{A}(t) = \begin{pmatrix}
        0 & g_{u_2}^*(t)g_{u_1}(t)& g_{u_2}^*(t)g_{v_1}(t)& g_{u_2}^*(t)g_{v_2}(t) \\
        -g_{u_1}^*(t)g_{u_2}(t) & 0 & g_{u_1}^*(t)g_{v_1}(t) & g_{u_1}^*(t)g_{v_2}(t) \\
        -g_{v_1}^*(t)g_{u_2}(t) & -g_{v_1}^*(t)g_{u_1}(t) & 0 & g_{v_1}^*(t)g_{v_2}(t) \\
        -g_{v_2}^*(t)g_{u_2}(t) & -g_{v_2}^*(t)g_{u_1}(t) & -g_{v_2}^*(t)g_{v_1}(t) & 0
    \end{pmatrix}.
\end{align}


\section{Cascaded System With Two Incident Coherent Gaussian Pulses}\label{sec:cascade}

We now solve the dynamics of this input-output theory model representing the SUPER scheme in order to analyze the underlying microscopic dynamics, which is reflected in the properties of the outgoing pulses. Specifically, we choose the physical parameters as used in the original SUPER paper~\cite{bracht_swing-up_2021}: a two-level system driven by two coherent, red-detuned and off-resonant Gaussian pulses. This regime requires intense laser pulses with large coherent-state amplitudes. Hence, we cannot use a full Hilbert space description of all modes, and we will approximate the input-output dynamics for the quantum pulses using a cumulant expansion approach~\cite{kubo_generalized_1962,plankensteiner_quantumcumulantsjl}, which proves reliable even in low order for initial coherent states. Details of the method and the numerical Julia code to derive and numerically solve the first-order cumulant equations are provided in Ref.~\cite{SUPER_example, zenodo_data}. 
 
To match the semiclassical coherent pulses in Eq.~(\ref{eq.SUPER}), we choose the input modes such that the coherent amplitude emitted by each virtual input cavity reproduces the corresponding classical drive. The Gaussian envelopes introduced in Sec.~\ref{sec:model} require the normalized mode functions

\begin{align}
    u_i(t) = \frac{1}{\sqrt{\sigma_i\sqrt{\pi}}}e^{-\frac{(t-\tau_i)^2}{2\sigma_i^2}}e^{-i\Delta_i t},
\end{align}
with the coherent-state amplitudes of the virtual input cavities $\alpha_i = \sqrt{n_{u_i}}$, where 
\begin{align}
     n_{u_i} = \frac{|A_i|^2}{8 \sqrt{\pi}\gamma\sigma_i}
    \label{eq.normal}
\end{align}
is the corresponding mean-photon number in the virtual input cavity $u_i$. 
Hence, for fixed pulse width $\sigma_i$ and pulse area $A_i$ (or peak Rabi frequency), the corresponding coherent amplitudes scale as $\alpha_i \propto \gamma^{-1/2}$ and the pulse occupations obey $n_{u_i}\propto 1/\gamma$. 
In the pulse description, the effective occupation of the incident fields is therefore set by the coupling strength (decay rate) $\gamma$ of the TLS to the radiation field. As we will see later, this leads to pulses with large mean photon numbers in order to excite the TLS well. If not explicitly stated otherwise, throughout this section, we choose $\gamma = 0.01~\mathrm{THz}$. The pulse parameters are listed in Tab.~\ref{tab.1}. We want to emphasize again that the peak Rabi frequency of the pulses is fully determined by the pulse area and the pulse width, as can be seen in Eq.~\eqref{eq.SUPER}. In order to relate our work with reference~\cite{bracht_swing-up_2021}, we use the same pulse parameters but vary the pulse coupling strength (decay rate) $\gamma$, which changes the corresponding coherent state amplitude and mean-photon number of the pulse. 

\begin{table}
\caption{\label{tab.1}
SUPER parameters in {SI-units}.}
\begin{ruledtabular}
\begin{tabular}{ccccccc}
 $\text{Pulse}$ & $A_i$ &$\Delta_i~{(\mathrm{THz})}$ &  $\sigma_i~{(\mathrm{ps})}$ & $\tau_i~{(\mathrm{ps})}$ & $\alpha_i$ & $n_{u_i}$\\
\hline
$1$ & $22.65\pi$ & ${-2\pi1.934}$ & $2.40$ & $12.00$ & $121.98$ & $14879$\\ 
$2$ & $19.29\pi$ & ${-2\pi4.634}$ & $3.04$ & $11.27$ & $92.30$ & $8519$ 
\end{tabular}
\end{ruledtabular}
\end{table}

\begin{figure}
\centering
\includegraphics[trim=0.0cm 4.8cm 0.0cm 4.5cm,clip,width=0.99\linewidth]{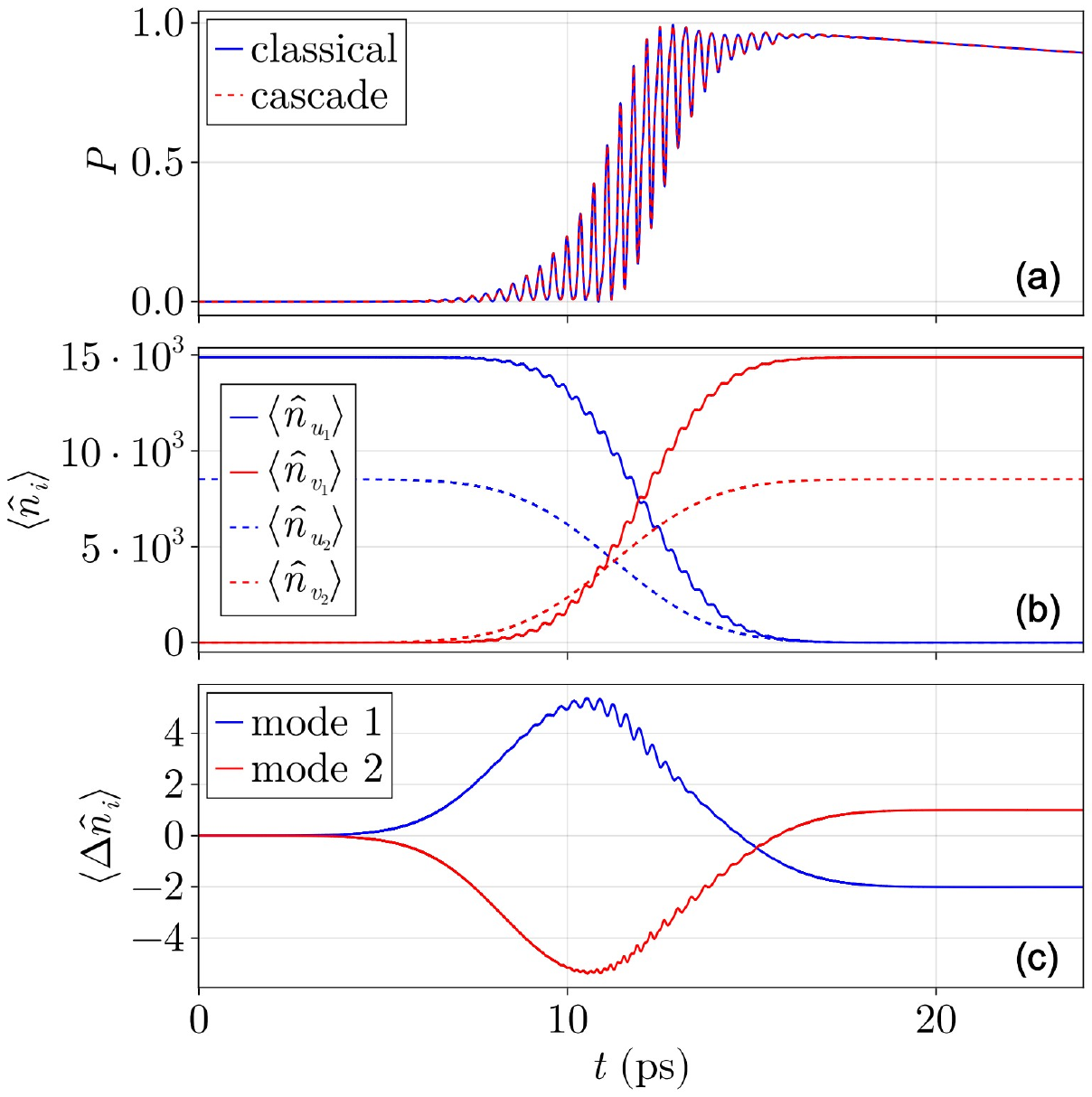}
\caption{Plot (a) demonstrates that the cascade input-output solution for the population $P = \langle \hat{\sigma}^+\hat{\sigma}^-\rangle$ coincides with the classical drive description. The mean photon numbers of each pulse $\langle\hat{n}_i\rangle = \langle \hat{a}_i^{\dagger}\hat{a}_i\rangle$ are depicted in plot (b). Initially, all photons are contained in the incident pulses $u_i(t)$ (blue lines) and the majority are transferred to the pulses $v_i(t)$ (red lines) during the excitation process. The final occupation number $\langle\hat{n}_{v_i}\rangle$ is changed slightly from the initial one $\langle\hat{n}_{u_i}\rangle$. Mode 1 experiences a loss of two photons (-2), whereas mode 2 gains one (+1), which is observed in plot (c).}\label{fig.res_cas1}
\end{figure}

The cascade's initial state is 

\begin{align}
    \ket{\psi_0}=\ket{\alpha_2}_{u_2}\otimes\ket{\alpha_1}_{u_1}\otimes\ket{g}_{\text{TLS}}\otimes\ket{0}_{v_1}\otimes\ket{0}_{v_2}.
\end{align} 
In Fig.~\ref{fig.res_cas1}, we simulate the dynamics of the system, where Fig.~\ref{fig.res_cas1}(a) visualizes that the input-output description reproduces the classical TLS population dynamics exactly, and Fig.~\ref{fig.res_cas1}(b) shows the corresponding photon dynamics in the virtual cavities. During the process, the input cavities are depleted as they emit the incident wave packets, while the output cavities are populated as they absorb the selected outgoing modes. To isolate the nontrivial scattering contribution, we define the photon loss/gain 
\begin{align}
    \langle \Delta \hat{n_i}(t)\rangle = \langle \hat{n}_{u_i}(t)\rangle + \langle \hat{n}_{v_i}(t)\rangle - \langle \hat{n}_{u_i}(0)\rangle.
\end{align}
This quantity describes the effective photon number change in the temporal mode $i\in\{1,2\}$. 

In Fig.~\ref{fig.res_cas1}(c), we see that at the end of the process, a net change of $-2$ photons in mode 1 and $+1$ photon in mode 2. This is the same characteristic $(-2,+1)$ signature found in the cavity realization of the SUPER scheme \cite{richter_few-photon_2025} and demonstrates that the traveling-pulse description retains the same multiphoton dynamics.

\subsection{Interaction-Picture Description}\label{se.int}

\begin{figure}
\centering
\includegraphics[width=0.90\linewidth]{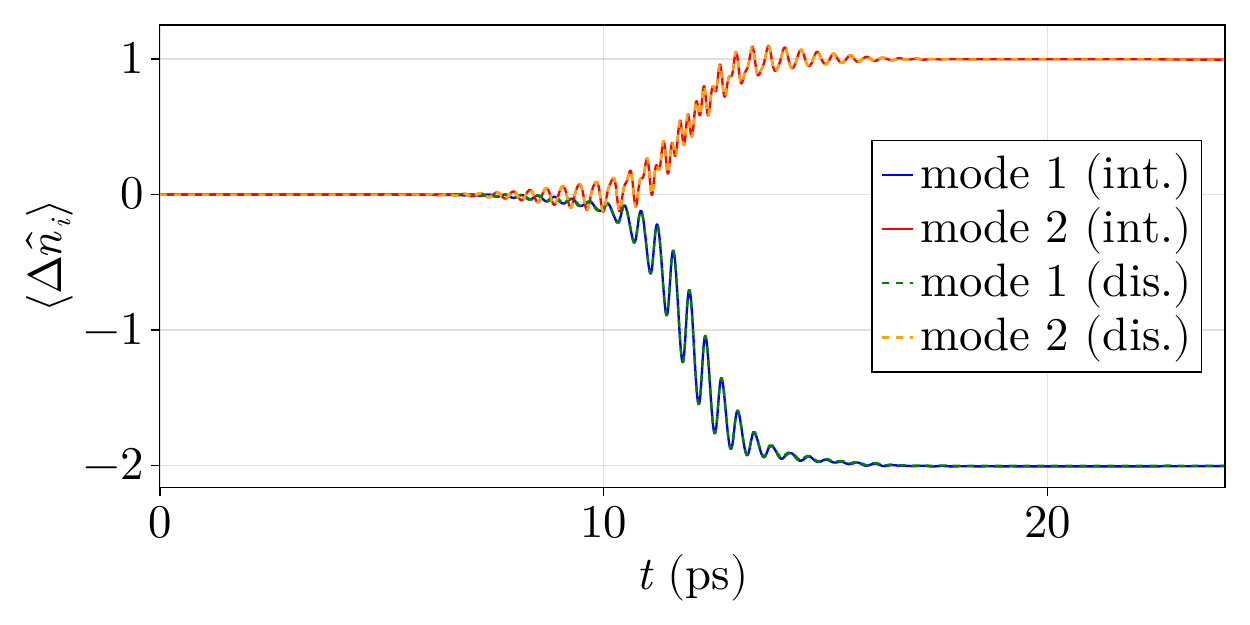}
\caption{
The photon loss/gain of each mode $\langle\Delta\hat{n}_i\rangle$ in the interaction picture (int.) remains the same, where the trajectories converge towards (-2,+1) directly. Applying the displacement frame (dis.) and solving the full master equation yields the exact same results as the cumulative interaction picture.}
\label{fig.res_dis}
\end{figure}

To remove the numerical effort involving the coherent transfer of unaffected photons from the input mode $u_i$ to the output mode $v_i$ and to isolate the small amplitude modifications of the field dynamics associated with the nontrivial scattering event, we transform our Hamiltonian $\hat{H}^{\text{cas}}(t)$ in Eq.~(\ref{eq.Ham}) into an interaction picture with respect to the input-output cavity transfer terms~\cite{christiansen_interactions_2023}, described by the Hamiltonian part 
\begin{align}
    \frac{i}{2}\boldsymbol{a}^{\dagger}\cdot\boldsymbol{A}(t)\cdot\boldsymbol{a}. 
\end{align}
The interaction-picture mode operators obey~\cite{christiansen_interactions_2023}
\begin{align}
    \frac{d}{dt}\hat{a}_{I,i}(t) = -i[\hat{a}_{I,i}(t),\hat{H}_0(t)]
\end{align}
with $i\in\{u_2,u_1,v_1,v_2\}$. As shown in Appendix~\ref{app.int}, the interaction-mode operators evolve as
\begin{align}
    \boldsymbol{a}_I(t) = \boldsymbol{M}(t)\boldsymbol{a}_I(0) = \boldsymbol{M}(t)\boldsymbol{a},
\end{align}
and the matrix $\boldsymbol{M}(t)$ satisfies
\begin{align}
    \frac{d}{dt}\boldsymbol{M}(t) = \frac{1}{2}\boldsymbol{A}(t)\boldsymbol{M}(t). 
\end{align}
The transformed Hamiltonian and jump operator are therefore

\begin{align}
    \hat{H}_{I}(t) &= \hat{H}_{\text{TLS}}- \frac{\sqrt{\gamma}}{2i}(\hat{\sigma}^+\tilde{\boldsymbol{g}}^{\dagger}(t)\boldsymbol{M}(t)\boldsymbol{a} - \text{h.c.})
\end{align}
and
\begin{align}
    \hat{L}_{I}(t) = \sqrt{\gamma}\hat{\sigma}^- + \boldsymbol{g}^{\dagger}\cdot \boldsymbol{M}(t)\cdot \boldsymbol{a}.
\end{align}
The interaction-picture results are shown in Fig.~\ref{fig.res_dis}. As in the previous cascade, the two-level dynamics are reproduced exactly (not shown), while the mode dynamics again show a net change of $-2$ and $+1$ photons. In this representation, however, the unaffected transfer between input and output cavities is removed, and the dynamics are carried only by the few photons that mediate the actual excitation process. This makes the physical origin of the SUPER transition more transparent and reduces the numerical complexity. In Sec.~\ref{sec:quantum}, we use this formulation as the basis for full master-equation calculations. 

\subsection{Minimum photon number for pulses}\label{sec:min} 
In Refs.~\cite{richter_few-photon_2025, Vannucci2024}, the SUPER scheme is analyzed with a two-mode Jaynes-Cummings model. Within this description, it is possible to fully invert the TLS with the initial state $\ket{g,2,0}$, i.e.\ even with zero photons in the second cavity mode. Due to the multimode continuum of eigenmodes, the traveling pulse description is different from the two-mode Jaynes-Cummings model~\cite{rueskov_christiansen_jaynescummings_2024}. This becomes particularly clear for the above case with zero photons in one mode. While the vacuum field of a cavity mode can strongly influence the emitter dynamics, the vacuum contribution of traveling-wave modes is incorporated through the spontaneous-emission Lindblad operator describing coupling to all modes orthogonal to the selected pulse mode. This suggests that the SUPER scheme described by traveling pulses requires a minimal number of photons, even if all the light is perfectly coupled to the TLS~\cite{Stobiska2009, Wang2011}. Furthermore, due to the underlying off-resonant three-photon process~\cite{junyu2025coherent} of the scheme, we expect that a photon number much larger than one is required to create a high excited state population. 

\begin{figure}
\centering
\includegraphics[trim=0.0cm 1.0cm 0.0cm 1.0cm,clip,width=0.99\linewidth]{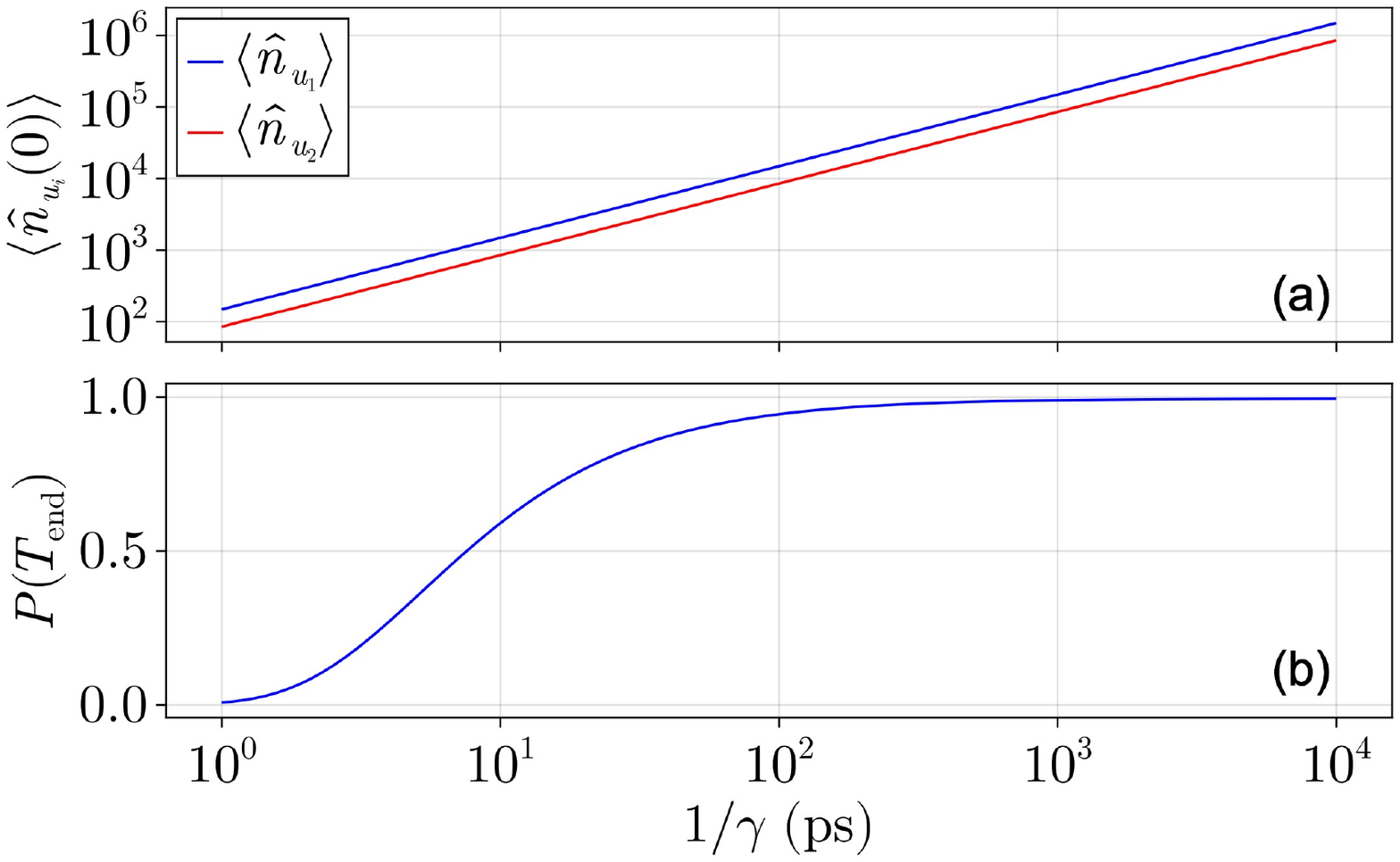}
\caption{In plot (a) the initial mean photon numbers $\langle\hat{n}_{u_i}(0)\rangle(1/\gamma)$ with $1/\gamma$ in logarithmic scaling are shown. Reducing the decay rate $\gamma$ implies a linear increase of $\langle \hat{n}_{u_i}(0)\rangle \propto 1/\gamma$. The final TLS population $P = \langle \hat{\sigma}^+\hat{\sigma}^- (T_{\text{end}})\rangle(1/\gamma)$ is depicted in plot (b). Here, $T_{\text{end}}$ denotes the time immediately after the SUPER excitation process is over.}\label{fig.photon}
\end{figure}

In FIG.~\ref{fig.photon} we show the maximum excited state population depending on the photon numbers in the pulses. As we keep the corresponding Rabi frequency pulse the same, we effectively change the photon number in both pulses simultaneously by varying the coupling rate $\gamma$, see Eq.~\eqref{eq.normal}. We see that to achieve an excited state population of above $99\%$, a couple of tens of thousands of photons in each mode are required. While reducing $\gamma$ leads to a population $P(T_{\text{end}})$ convergence towards $1$, increasing the said TLS rate results in a population decrease, leading eventually to no excitation at all. We also want to note that the studied system is idealized such that the traveling pulses couple perfectly to the TLS, corresponding to an excitation via an ideal chiral waveguide. 

One might suggest increasing the interaction strength of the traveling pulse by reducing its temporal width; however, this leads to the time-bandwidth dilemma~\cite{kiilerich_quantum_2020}: if the pulse gets shorter, assuming a modification keeping the initial mean photon numbers constant, most of it will be off-resonant with the desired frequency detuning. While there might be an optimal pulse width and detuning, which potentially allows for fewer photons to achieve a high population, this will not substantially change the behavior. To verify this, we perform a numerical optimization to find the parameters with the smallest mean photon number that achieve a final state population of $P(T_{\text{end}}) > 80 \%$. The smallest mean-photon number we found was still above 2000, see Appendix~\ref{app.minimum}. This confirms that the off-resonant multi-photon SUPER process requires a large number of photons for traveling pulses. To observe the process with a few photons, a cavity interaction in the strong-coupling regime is needed~\cite{richter_few-photon_2025, Vannucci2024}. 


\section{Full Quantum Description} \label{sec:quantum}
The cumulant expansion provides an efficient description of the large-photon-number coherent-pulse dynamics, but its accuracy might still be doubted. Fortunately, the interaction-picture formulation also allows direct full-quantum simulations in a modified reference frame for the virtual cavities. In this section, we first show how the previous dynamics can be solved in a suitably chosen displacement frame, where the large coherent amplitudes are treated as quantum fluctuations from their classical mean-field background. In this limit, the quantum fluctuations remain small enough to allow for explicit full representation. In a second step, we then also turn to the case of genuine quantum input pulses starting from fixed photon number initial Fock states, which requires a full quantum description.

\subsection{Displacement Frame Description}
Although the average photon number does not significantly change in the interaction picture formulation, the corresponding truncated Hilbert space still needs to be very large as it has to cover the important range of the photon number distribution of the coherent states, which exhibits an uncertainty proportional to the square root of the mean photon number. It is thus advantageous to transform the interaction-picture Hamiltonian into a displacement frame~\cite{glauber_coherent_and_incoherent,gardiner_input_and_output,carmichael_open_1993,gardiner_quantum_noise, Wiseman_Milburn_2009}, defined by the unitary operator
\begin{align}
    \hat{\mathfrak{D}} = \prod_{j\in\{1,2\}} \hat{\mathcal{D}}_{u_j}(\alpha_{u_j}),
\end{align}
where $\hat{\mathcal{D}}_{u_j}(\alpha_{u_j}) = \exp({\alpha_{u_j}\hat{a}_{u_j}^{\dagger} - \alpha^*_{u_j}\hat{a}_{u_j}})$ is the displacement operator associated with the pulse $u_j$. As this only adds a c-number to the field mode operators, their commutation relations remain unchanged, and we obtain the transformation
\begin{align}
    \hat{\mathcal{D}}^{\dagger}_{u_j}(\alpha_{u_j})\hat{a}_{u_j}\hat{\mathcal{D}}_{u_j}(\alpha_{u_j}) = \hat{a}_{u_j} + \alpha_{u_j}\mathds{1}.
\end{align}
In this frame, the $c$-number amplitude $\alpha_j$ accounts for the coherent average amplitude and $\hat{a}_j$ describes the fields' quantum fluctuations. Transforming the interaction density matrix accordingly yields
\begin{align}
    \hat{\rho}_D = \hat{\mathfrak{D}}^{\dagger}\hat{\rho}_I\hat{\mathfrak{D}}.
\end{align}
The interaction Hamiltonian in the displacement frame then reads
\begin{align}
    \hat{H}_D(t) &= \hat{\mathfrak{D}}^{\dagger}\hat{H}_{I}(t)\hat{\mathfrak{D}} \underbrace{-i\hat{\mathfrak{D}}^{\dagger}\dot{\hat{\mathfrak{D}}}}_{=0}\\
    &= \hat{H}_{\text{TLS}} + \underbrace{\frac{i\sqrt{\gamma}}{2}(\hat{\sigma}^+\tilde{\boldsymbol{g}}^{\dagger}(t)\boldsymbol{M}(t)\boldsymbol{\alpha} - \text{h.c.})}_{\text{classical~drive}}\\
    &+  \underbrace{\frac{i\sqrt{\gamma}}{2}(\hat{\sigma}^+\tilde{\boldsymbol{g}}^{\dagger}(t)\boldsymbol{M}(t)\boldsymbol{a} - \text{h.c.})}_{\text{quantum-fluctuations}},
\end{align}
where $\boldsymbol{\alpha} = (\alpha_{u_2},\alpha_{u_1},0,0)^T$. The displacement frame thus separates the dynamics into a classical driving part $(\boldsymbol{\alpha})$ and a fluctuation part around the vacuum state $(\boldsymbol{a})$. Since the field operators are shifted by $c$-numbers, the jump operator obtains the form~\cite{glauber_coherent_and_incoherent,gardiner_input_and_output,carmichael_open_1993,gardiner_quantum_noise,Wiseman_Milburn_2009},
\begin{align}
    \hat{L}_D = \hat{\mathfrak{D}}^{\dagger}\hat{L}_{I}(t)\hat{\mathfrak{D}} =  \sqrt{\gamma}\hat{\sigma}^-+ \boldsymbol{g}^{\dagger}(t)\boldsymbol{M}(t)(\boldsymbol{a} + \boldsymbol{\alpha}).
\end{align}
The initial coherent two-mode product state then transforms to

\begin{align}
    \ket{\psi_0}\to\ket{\psi_0'} = \hat{\mathfrak{D}}^{\dagger}\ket{\psi_0} = \ket{0}_{u_2}\ket{0}_{u_1}\ket{g}_{\text{TLS}}\ket{0}_{v_1}\ket{0}_{v_2},
\end{align}
which drastically reduces the required Hilbert-space dimension and permits a full numerical solution of the master equation with a reasonable cut-off. The resulting dynamics is shown in FIG.~\ref{fig.res_dis} and the average values and variances exhibit very good agreement with the cumulant-based interaction-picture solution from above.  

\subsection{Pure Fock State Description}\label{sec.fock}
\begin{figure}
    \centering
    \includegraphics[trim=0.0cm 1.5cm 0.0cm 1.0cm,clip,width=0.99\linewidth]{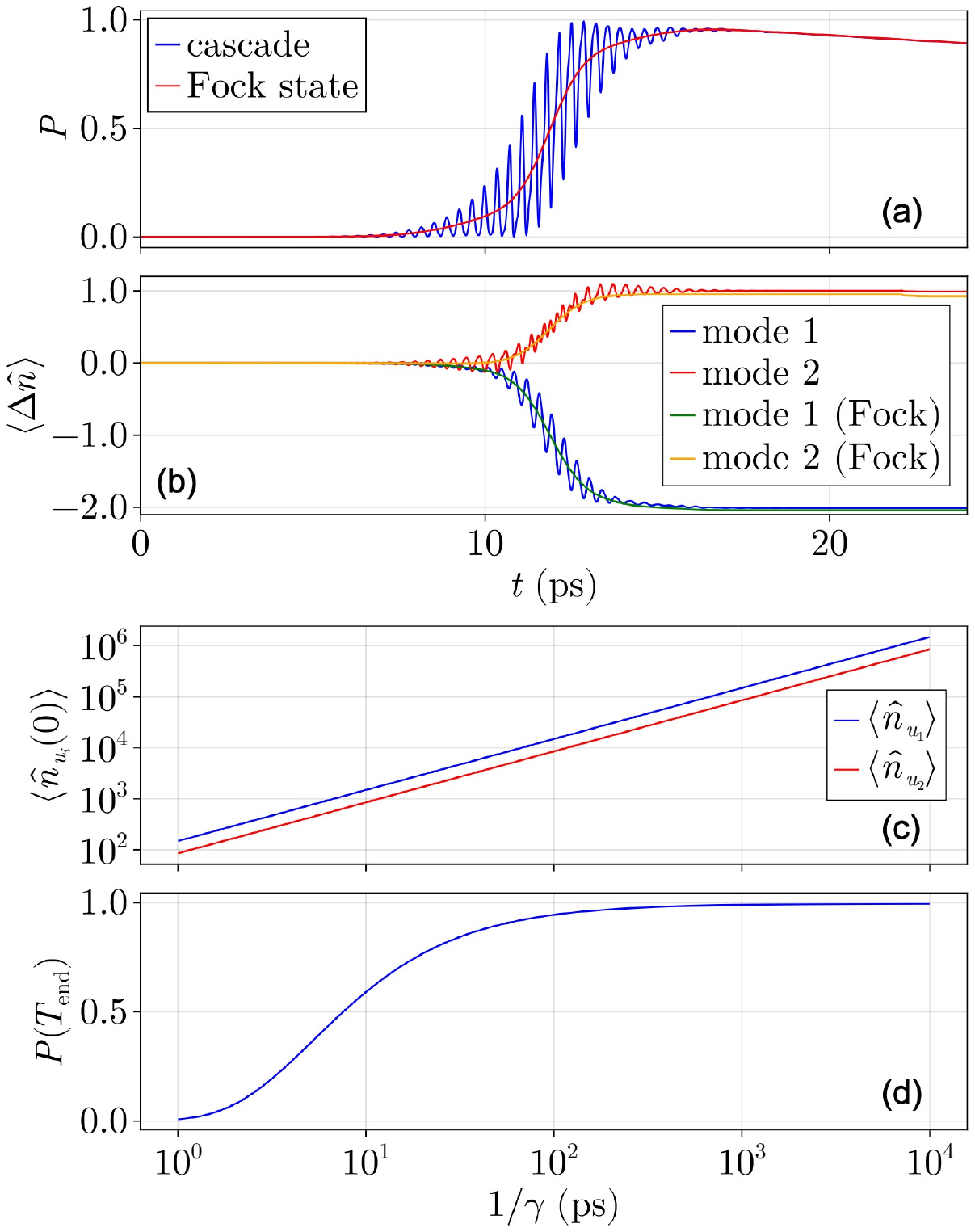}
    \caption{The pure Fock state population dynamics $P$ is compared to the cascaded system in plot (a). The fundamental excitation process without additional oscillatory behavior can be observed clearly. Similar is the case for the loss/gain functions $\langle \Delta\hat{n}_i\rangle$ in plot~(b), which fit almost perfectly. Due to the choice of time steps and size of the truncated Fock spaces, small deviations appear. In plot (c) and (d), we visualize that the success of the SUPER scheme in the Fock space full-master equation description is subject to the same $1/\gamma$ dependence as shown previously. Since the fundamental process is the same, this behavior is expected.}\label{fig.res_fock}
\end{figure}

Let us now compare the dynamics for coherent input pulses with the case of incident non-classical photon number eigenstates, where we choose states with the same mean photon numbers as for the coherent pulses, i.e., $n_{u_1} = 14879$ and $n_{u_2} = 8519$ in Tab.~\ref{tab.1} and utilizing $\gamma = 0.01$. 
Since in the atom-field interaction only a few photons are exchanged, the quantum states of the excitation pulses are only changed by a couple of photons, we only need to keep a couple of nearby Fock states in the computational basis. A corresponding numerical solution of the master equation for a suitably truncated Hilbert space in Eq.~(\ref{eq.fock}) is shown in Fig.~\ref{fig.res_fock}(a)-(b). Interestingly, the non-classical Fock-state input based simulation reproduces the qualitative behavior of the cumulant-based solution before, and, in particular, it exhibits the same photon-exchange pattern between the modes. As a major qualitative difference to the coherent-state case, however, the time evolution is much smoother with the fast oscillations absent. This originates from the undefined phase of the Fock state, which prevents interference. For one pulse in a coherent state and the other in a Fock state, the fast oscillations are also absent (not shown), confirming our explanation. 
Hence, although the field phase for Fock states is undefined, the atom follows the same overall swing-up dynamics as for coherent incoming pulses, but in a smoother form.

Since the dynamics are otherwise very similar, we also expect the same qualitative threshold behavior with respect to the pulse occupation. This is indeed what we find in Fig.~\ref{fig.res_fock}(c)-(d): Fock-state pulses also require sufficiently large photon numbers to achieve a high excited-state population. In fact, we observe the same $\langle \hat{\sigma}^{+}\hat{\sigma}^-(T_{\text{end}})\rangle(1/\gamma)$ curve as in Fig.~\ref{fig.photon}. 
Unfortunately, for lower photon numbers in the two input modes, the achievable final inversion fidelity drops as well. This shows that the minimum-photon-number requirement is a generic feature of the pulsed non-linear SUPER scheme and not simply a consequence of the reduced coherent-state number fluctuations at larger amplitudes. 

\section{Conclusion}
In summary, we presented a detailed analysis of the two-color SUPER quantum emitter excitation scheme within the input-output theory for quantum pulses. We used this framework to analyze both the emitter dynamics and the accompanying pulse-mode dynamics. The central result is that the traveling-pulse realization of SUPER exhibits the same multi-photon transition character previously identified in cavity-based model simulations. In our free-space formulation, the input-output treatment makes the three-photon Raman transition structure directly visible in the time-dependent pulse occupation. In particular, the mode-resolved dynamics reveal the characteristic photon-number change of $-2$ in one mode and $+1$ in the other, providing a direct pulse-based signature of the underlying multiphoton process. At the same time, our analysis shows that the traveling-pulse problem is not equivalent to a two-mode cavity description: for fixed pulse shapes, high-fidelity inversion requires a large pulse occupation, which cannot be circumvented for traveling pulses.

Beyond the specific physical example studied, we used the generic SUPER scheme to showcase several alternative approaches for numerical treatment of quantum light pulse dynamics. By combining the pulse-based input-output description with the cumulant expansion, we reproduced the classical two-level dynamics of the original SUPER protocol while retaining access to the field degrees of freedom. The interaction-picture transformation showed that the nontrivial dynamics are governed only by a few exchanged photons, which in turn enables direct full-quantum calculations via a displacement-frame treatment or for incident pulses in large Fock states with definite photon number. These methods should be useful more broadly for further investigations of excitation, scattering, and state-transfer problems in which the temporal mode structure of the fields is an essential part of the physics.

\acknowledgements 
This research was funded in whole or in part by the Austrian Science Fund (FWF) projects 10.55776/FG5 (Forschungsgruppe FG 5), the quantA cluster of excellence 10.55776/COE1, and 10.55776/J4865. The author has applied a CC BY public copyright license to any Author Accepted Manuscript (AAM) version arising from this submission. C.H. was supported by the Carlsberg Foundation through the “Semper Ardens” Research Project QCooL. The data presented in this article are available from Ref.~\cite{zenodo_data}.


\bibliography{references}

\newpage
\clearpage


\appendix

\section{Multi-Mode Pulse Couplings}\label{app:couplings}
For a cascade with several input modes $\{u_i(t)\}_{i=1}^{m}$ and several selected output modes $\{v_i(t)\}_{i=1}^{n}$, the effective virtual-cavity couplings depend on the ordering of the cascade. This reflects the fact that a mode emitted by one virtual cavity is modified by the downstream cavities before reaching the localized system, and similarly for the selected output modes.

For the input side, the couplings take the form~\cite{kiilerich_input-output_2019,kiilerich_quantum_2020,christiansen_interactions_2023} 

\begin{align}
    g_{u_i}(t) = \frac{[u_i^{(i-1)}(t)]^*}{\sqrt{1 - \int_0^{t}dt'|u_i^{(i-1)}(t')|^2}},
\end{align}
where the effective mode emitted by the $i$-th cavity obeys
\begin{align}
    u_i^{(i-1)}(t) = u_i(t) - \sum_{k=1}^{i-1} g_{u_k}^*(t)\alpha_{u_i}^{(k)}(t),
\end{align}
and the auxiliary (distortion) amplitudes satisfy
\begin{align}
    \partial_t\alpha_{u_i}^{(j)}(t) =
    -g_{u_j}(t)\left[u_i(t) - \sum_{k=1}^{j-1}g_{u_k}^*(t)\alpha_{u_i}^{(k)}(t)\right]
    + \frac{|g_{u_j}(t)|^2}{2}\alpha_{u_i}^{(j)}(t).
\end{align}
The initial conditions are $\alpha_{u_i}^{(j)}(0)=0$.

For the selected output modes, the corresponding expressions are
\begin{align}
    g_{v_i}(t) &= -\frac{[v_i^{(i-1)}(t)]^*}{\sqrt{\int_0^{t}dt'|v_i^{(i-1)}(t')|^2}}
\end{align}
\begin{align}
    v_i^{(i-1)}(t) = v_i(t) + \sum_{k=1}^{i-1} g_{v_k}^*(t)\alpha_{v_i}^{(k)}(t)
\end{align}
\begin{align}
    \partial_t\alpha_{v_i}^{(j)}(t) =
    -g_{v_j}(t)\left[v_i(t) + \sum_{k=1}^{j-1}g_{v_k}^*(t)\alpha_{v_i}^{(k)}(t)\right]
    - \frac{|g_{v_j}(t)|^2}{2}\alpha_{v_i}^{(j)}(t),
\end{align}
again with $\alpha_{v_i}^{(j)}(0)=0$. These relations are the multi-mode generalization of the single-mode couplings used in the main text.

\section{Minimal mean-photon number parameter optimization}\label{app.minimum}

We ran a numerical optimization to minimize the total mean-photon number $n_{u_1} + n_{u_2}$ of the pulses, which achieves a final excited state population above $80 \%$. The following restrictions are used in the optimization: 
The detunings need to be negative and $|\Delta_2| > 2 |\Delta_1|$, to fulfill the SUPER condition~\cite{bracht_swing-up_2021, richter_few-photon_2025}. To avoid near-resonant excitation, we restrict the Rabi frequencies to $\Omega_i < \Delta_i$. The optimization parameters are restricted to the following ranges: 
\begin{itemize}
    \item $0 < A_1, A_2 < 50\pi$ 
    \item $0.1~\mathrm{ps} < \sigma_1, \sigma_2 < 10~\mathrm{ps}$ 
    \item $-2\pi 10~\mathrm{THz} < \Delta_1, \Delta_2 < -2\pi 0.5~\mathrm{THz}$ 
    \item $-5~\mathrm{ps} < \tau < 5~\mathrm{ps}$
     \item  $0.001~\mathrm{THz} < \gamma < 1~\mathrm{THz}$
\end{itemize}

This resulted in the set of parameters $A_1 = 24.944, \sigma_1 = 1.1286~\mathrm{ps}, \Delta_1 = -4.4090~\mathrm{THz}, A_2 = 21.704, \sigma_2 = 1.1306~\mathrm{ps}, \Delta_2 = -14.007~\mathrm{THz}, \tau = -0.01215~\mathrm{ps}, \gamma = 0.03271~\mathrm{THz}$ with a total mean-photon number of $2086.7$ ($n_{u_1} = 1188.5$ and  $n_{u_2} = 898.2$). The optimization program can be found in Ref.~\cite{zenodo_data}. 


\section{Interaction-Picture Construction}\label{app.int}

The interaction Hamiltonian is given by

\begin{align}
    \hat{H}_0(t) = \frac{i}{2}\boldsymbol{a}^{\dagger}\boldsymbol{A}(t)\boldsymbol{a}
\end{align}
with the unitary operator
\begin{align}
    \hat{U}_0(t) = \mathcal{T}\exp\bigg\{-i\int_0^{t}dt'\hat{H}_0(t')\bigg\}.
\end{align}
for the time evolution of the interaction picture operators
\begin{align}
    \boldsymbol{a}_I(t) = \hat{U}^{\dagger}_0(t)\boldsymbol{a}\hat{U}_0(t).
\end{align}
The Schr\"odinger equation provides:

\begin{align}
    \dot{\hat{U}}_0(t) &= -i\hat{H}_0(t)\hat{U}_0(t)\\
    \dot{\hat{U}}^{\dagger}_0(t) &= i\hat{U}^{\dagger}_0(t)\hat{H}_0(t)
\end{align}
which can be inserted into

\begin{align}
    \dot{\boldsymbol{a}}_I(t) &= \dot{\hat{U}}^{\dagger}_0(t)\boldsymbol{a}\hat{U}_0(t) + \hat{U}^{\dagger}_0(t)\boldsymbol{a}\dot{\hat{U}}_0(t)\\
    &= i\hat{U}^{\dagger}_0(t)[\hat{H}_0(t),\boldsymbol{a}]\hat{U}_0(t) \\
    &= -\frac{1}{2}\hat{U}^{\dagger}_0(t)[\boldsymbol{a}^{\dagger}\boldsymbol{A}(t)\boldsymbol{a},\boldsymbol{a}]\hat{U}_0(t) \\
    &=-\frac{1}{2}\sum_k\sum_{ij}A_{ij}(t)\hat{U}^{\dagger}_0(t)\underbrace{[\hat{a}_i^{\dagger}\hat{a}_j,\hat{a}_k]}_{-\delta_{ik}\hat{a}_j}\boldsymbol{\varepsilon}_k\hat{U}_0(t) \\
    &=\frac{1}{2}\sum_{kj}A_{kj}(t)\hat{U}^{\dagger}_0(t)\hat{a}_j\boldsymbol{\varepsilon}_k\hat{U}_0(t)\\
    &=\frac{1}{2}\boldsymbol{A}(t)\hat{U}^{\dagger}_0(t)\boldsymbol{a}\hat{U}_0(t) = \frac{1}{2}\boldsymbol{A}(t)\boldsymbol{a}_I(t).
\end{align}
Solving the above equation yields

\begin{align}
    \boldsymbol{a}_I(t) = \boldsymbol{M}(t)\boldsymbol{a}_I(0) = \boldsymbol{M}(t)\boldsymbol{a},
\end{align}
where 

\begin{align}
    \boldsymbol{M}(t) = \mathcal{T}\exp\bigg\{\frac{1}{2}\int_{0}^{t}dt'\boldsymbol{A}(t')\bigg\}.
\end{align}

\section{Fock Space Truncation}
For the Fock space truncation in Sec. \ref{sec.fock}, we consider the Fock space for bosonic particles 
\begin{align}
    \mathfrak{F} = \bigoplus_{n = 0}^{\infty}\mathcal{S}\mathcal{H}^{\otimes n},
\end{align}
where $\mathcal{S}$ is the bosonic symmetry operator and $\mathcal{H}$ denotes the single-particle Hilbert space. The individual pulse Fock spaces are truncated such that they take on the following form
\begin{align}
    \mathfrak{F}_{u_j} = \bigoplus_{n = n_{u_j} - N^l_{u_j}}^{n_{u_j}+N^u_{u_j}}\mathcal{S}\mathcal{H}^{\otimes n}.&&N^l_{u_j},N^u_{u_j}\in\mathbb{N}\label{eq.fock}
\end{align}
For the $1/\gamma$-sweep Fig.~\ref{fig.res_fock}(c) and Fig.~\ref{fig.res_fock}(d), we used a smaller truncation of the Hilbert space than for $P$ and $\langle \Delta \hat{n}_i \rangle$ in Fig.~\ref{fig.res_fock}(a) and Fig.~\ref{fig.res_fock}(b), respectively. Whereas the correct loss/gain curves $\langle \Delta\hat{n}_i\rangle$ emerge from larger truncated Fock spaces, e.g. $(N_{u_1}^l = 6, N_{u_1}^u = 2)$ and $(N_{u_2}^l = 3, N_{u_2}^u = 5)$, the correct $P$ dynamics are already provided for smaller spaces, e.g. $(N_{u_1}^l = 2, N_{u_1}^u = 1)$ and $(N_{u_2}^l = 1, N_{u_2}^u = 2)$. The first case is used for Fig.~\ref{fig.res_fock}(a)-(b) and the second one for Fig.~\ref{fig.res_fock}(c)-(d). 

\section{Concatenated Description of the Two-Pulse SUPER Scheme}\label{sec.concat}

\begin{figure}
\centering
\includegraphics[clip, trim=0.0cm 10.5cm 1.0cm 0.0cm, width=.99\linewidth]{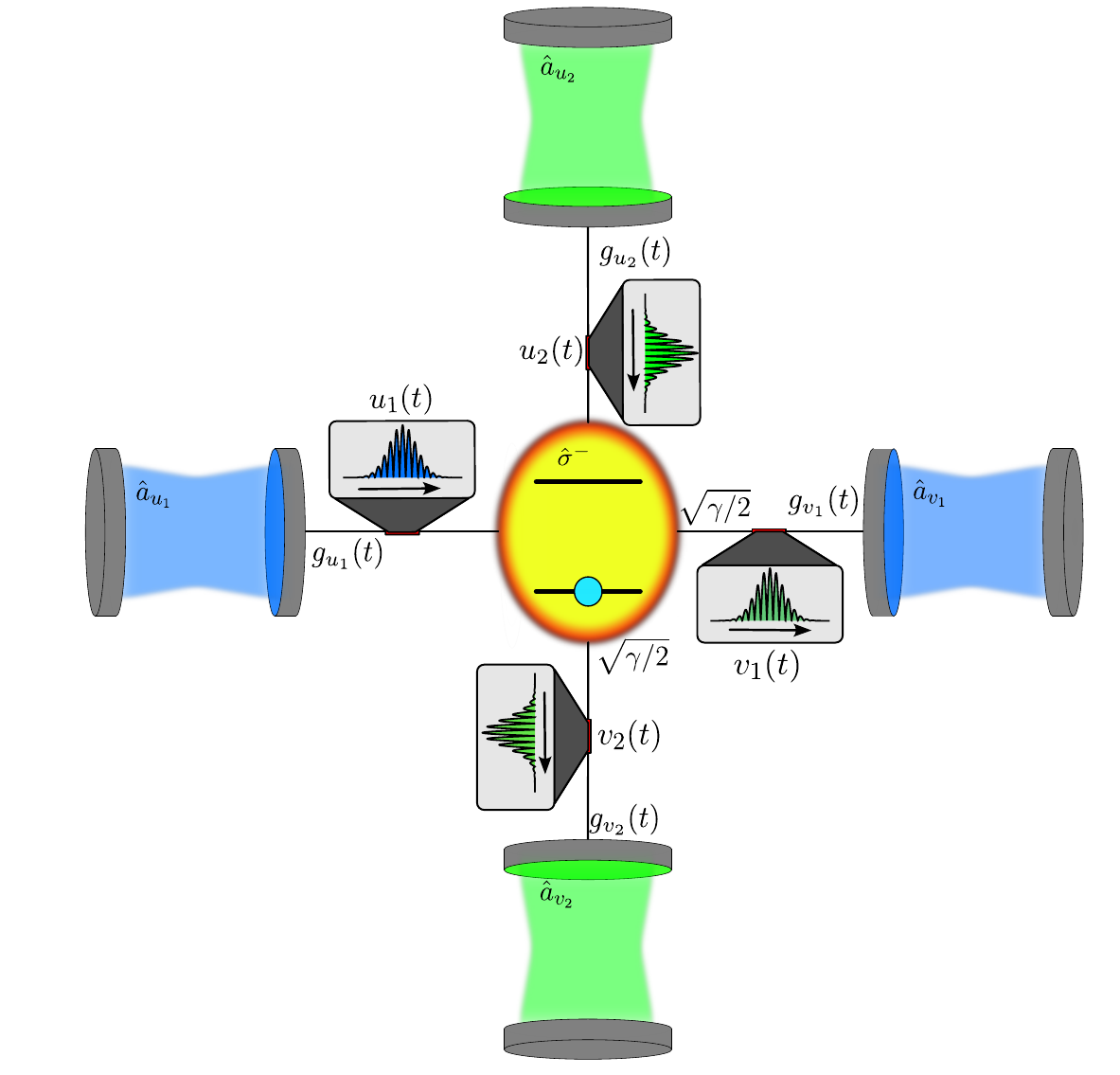}
\caption{Alternative description of the SUPER protocol as the concatenation of two pulse channels.}
\label{fig.schem_conc}
\end{figure}

The main text employs a single ordered cascade containing both input pulses. As an alternative formulation, one may assign each pulse to its own cascaded channel and combine the two channels by the SLH concatenation rule~\cite{combes_slh_nodate}
\begin{align}
    G_1 \boxplus G_2 =
    \left(
    \begin{pmatrix}
        \hat{S}_1 & 0 \\
        0 & \hat{S}_2
    \end{pmatrix},
    \begin{pmatrix}
        \hat{L}_1 \\
        \hat{L}_2
    \end{pmatrix},
    \hat{H}_1+\hat{H}_2
    \right), 
\end{align}
with $G_1=(\hat{S}_1,\hat{L}_1,\hat{H}_1)$ and $G_2=(\hat{S}_2,\hat{L}_2,\hat{H}_2)$. This construction describes two independently prepared pulses impinging on the same emitter as depicted in Fig.~\ref{fig.schem_conc}. 
For channel $i\in\{1,2\}$, we define
\begin{align}
    G_{\text{Ch},i} = G_{v_i}\lhd G_{\text{TLS}}\lhd G_{u_i},
\end{align}
and combine the two channels according to
\begin{align}
    G^{\text{conc}} = G_{\text{Ch},1}\boxplus G_{\text{Ch},2}
    = (\hat{S}^{\text{conc}},\hat{L}^{\text{conc}},\hat{H}^{\text{conc}}).
\end{align}
The channel operators are
\begin{align}
    \hat{L}_{\text{Ch},i} &= \sqrt{\frac{\gamma}{2}}\hat{\sigma}^- + \boldsymbol{g}_i^{\dagger}(t)\boldsymbol{a}_i
\end{align}
\begin{align}
    \hat{H}_{\text{Ch},i} &= \frac{\hat{H}_{\text{TLS}}}{2}
    + \frac{i}{2}\boldsymbol{a}_{i}^{\dagger}\boldsymbol{A}_i(t)\boldsymbol{a}_{i}
    - \frac{\sqrt{\gamma}}{2\sqrt{2}i}\left(\hat{\sigma}^+\tilde{\boldsymbol{g}}_i^{\dagger}(t)\boldsymbol{a}_i - \text{h.c.}\right),
\end{align}
where $\boldsymbol{a}_{i} = (\hat{a}_{u_i},\hat{a}_{v_i})^T$, $\boldsymbol{g}_i=(g^*_{u_i},g^*_{v_i})^T$, $\tilde{\boldsymbol{g}}_i=(g^*_{u_i},-g^*_{v_i})^T$, and
\begin{align}
  \boldsymbol{A}_i(t) = \begin{pmatrix}
      0 & g_{u_i}^*(t)g_{v_i}(t) \\
      -g_{u_i}(t)g_{v_i}^*(t) & 0
  \end{pmatrix}.
\end{align}
The factors of $1/2$ and $1/\sqrt{2}$ ensure that the TLS Hamiltonian and decay are not double counted when both channels are combined.

\begin{figure}
\centering
\includegraphics[trim=0.0cm 3.5cm 0.0cm 3.0cm,clip,width=0.99\linewidth]{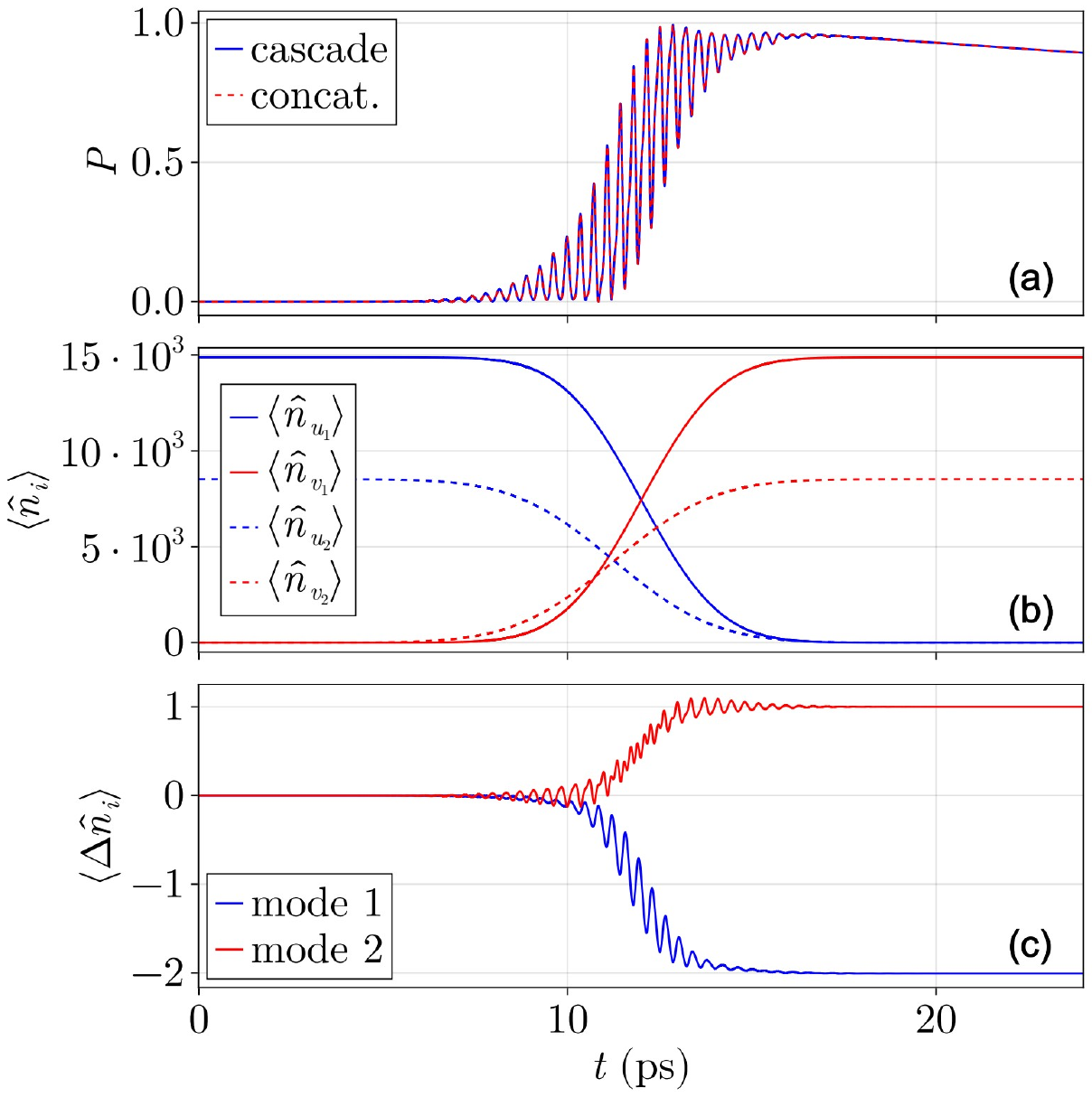}
\caption{Plot (a) visualizes the correct population $P$ dynamics obtained by the concatenated input-output SUPER scheme. In plot (b), the same photon exchange between the input and the output pulses as in the cascaded case can be observed, where the curves of $\langle \hat{n}_i\rangle$ contain no oscillations. This can be traced back to the fact that the input and output cavities only couple to each of the same mode. As a result, we observe the exact same loss/gain curves $\langle \Delta \hat{n} \rangle$ in plot (c) as obtained via the interaction picture.}
\label{fig.res_conc}
\end{figure}

Fig.~\ref{fig.res_conc}(a) shows that the concatenated network reproduces the same population inversion as the ordered cascade. The global photon transfer between incoming and outgoing modes is likewise unchanged. The main visible difference is that the small oscillatory features present in the ordered cascade are absent here, which reflects the fact that the two channels are treated independently. The concatenated description provides an instructive alternative representation of the same physical process.

\section{Single-Pulse Representation of the SUPER Drive}

Instead of treating the SUPER drive as two separate coherent pulses, one may also describe the full classical field $\Omega_{\text{S}}(t)$ of Eq.~(\ref{eq.SUPER}) as a single coherent pulse occupying one temporal mode. This representation reproduces the same emitter dynamics as the two-pulse description, but it no longer resolves how photons are exchanged between the two spectral components of the drive. Although the internal photon-dynamics remain mainly hidden, this description reduces the Hilbert space complexity and numerically accelerates other investigation approaches such as correlation-based modes determination utilizing the $g^{(1)}(t_1,t_2)$ auto-correlation function \cite{kiilerich_quantum_2020,lund_perfect_2023}. In~\cite{zenodo_data}, we calculate the temporal mode of the emitted single photon after the excitation.  

\begin{figure}
\centering
\includegraphics[trim=0.0cm 3.5cm 0.0cm 3.0cm,clip,width=0.99\linewidth]{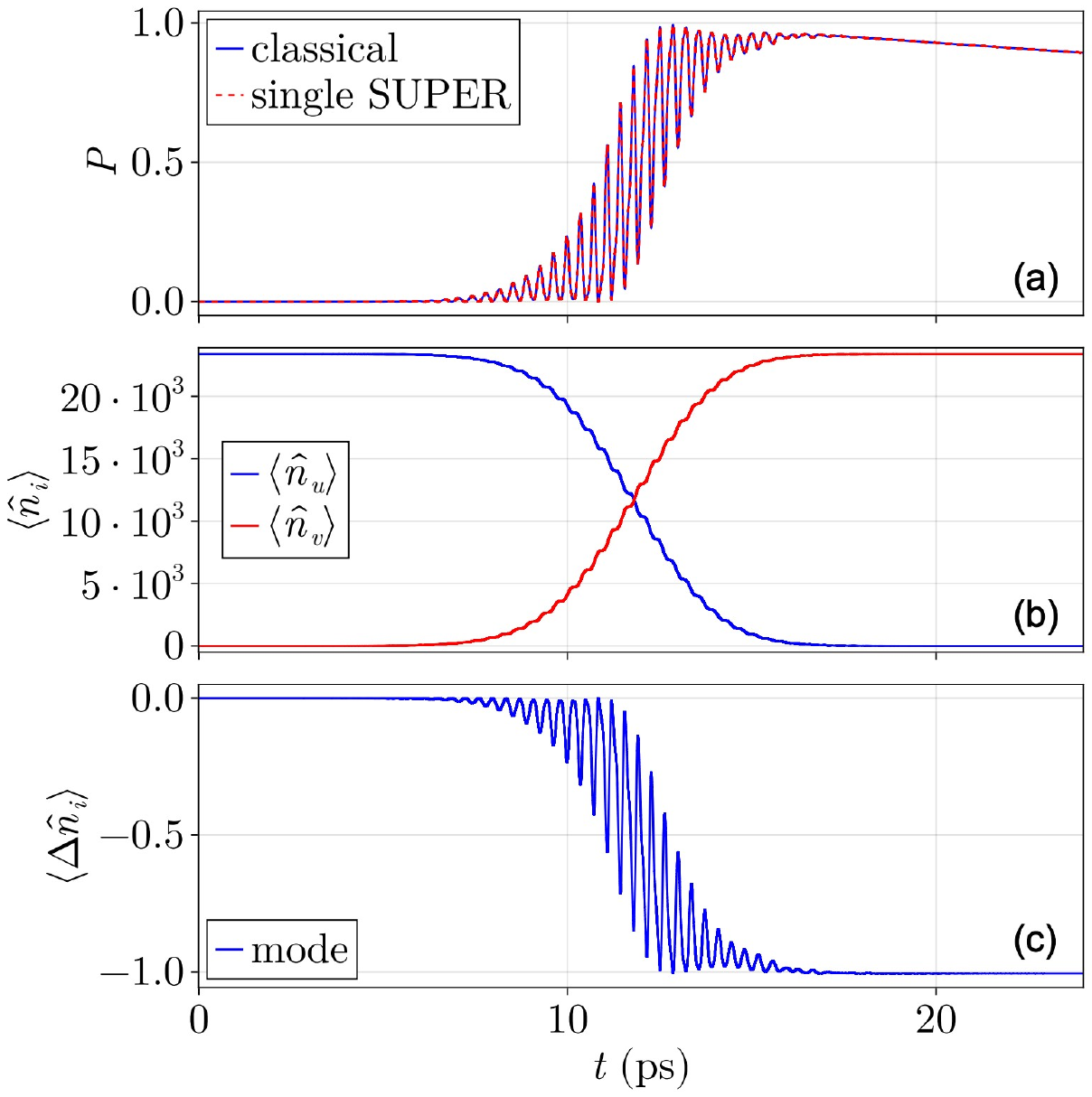}
\caption{The population $P$ of the single SUPER pulse cascade coincides perfectly with the classical dynamics as shown in plot (a). In plot (b), we find that $\langle \hat{n}_u \rangle$ of the input cavity is depleted and the majority of the photons are transferred into the output cavity. In fact, since we only obtain the net loss curve $\langle\Delta\hat{n}\rangle$ of the globally lost photon in plot (c), all photons but one are transferred from the input into the output pulse.}
\label{fig.res_cas2}
\end{figure}

To match the semiclassical SUPER pulse, we choose the single input mode
\begin{align}
    u(t)=\frac{\Omega_{\text{S}}(t)}{\sqrt{\int_0^{\infty} dt\,|\Omega_{\text{S}}(t)|^2}},
\end{align}
with coherent-state amplitude of the corresponding virtual input cavity
\begin{align}
    \alpha = \frac{1}{2\sqrt{\gamma}}\sqrt{\int_0^{\infty} dt\,|\Omega_{\text{S}}(t)|^2}.\label{eq.alpha_single}
\end{align}
For the Gaussian two-color pulse introduced in Sec.~\ref{sec:model}, the normalization integral is
\begin{align}
    \int_{0\to-\infty}^{\infty} dt\,|\Omega_{\text{S}}(t)|^2
    = \frac{A_1^2}{2\sqrt{\pi}\sigma_1}
    + \frac{A_2^2}{2\sqrt{\pi}\sigma_2}
    + 2\mathfrak{Re}\{S\},
\end{align}
where
\begin{align}
    S &= \frac{A_1A_2}{\sqrt{2\pi(\sigma_1^2 + \sigma_2^2)}}\exp\bigg(-\frac{(\tau_1 - \tau_2)^2}{2(\sigma_1^2 + \sigma_2^2)} \nonumber\\
    &\qquad - \frac{(\Delta_1 - \Delta_2)^2\sigma_1^2\sigma_2^2}{2(\sigma_1^2 + \sigma_2^2)} + i(\Delta_1 - \Delta_2)\frac{\sigma_1^2\tau_2 + \sigma_2^2\tau_1}{\sigma_1^2 + \sigma_2^2}\bigg).
\end{align}
With the parameters of Tab.~\ref{tab.1} the additional term $S$ is negligible.
\begin{figure}[!h]
\centering
\includegraphics[clip, trim=2.5cm 20.5cm 5.0cm 2.0cm, width=.99\linewidth]{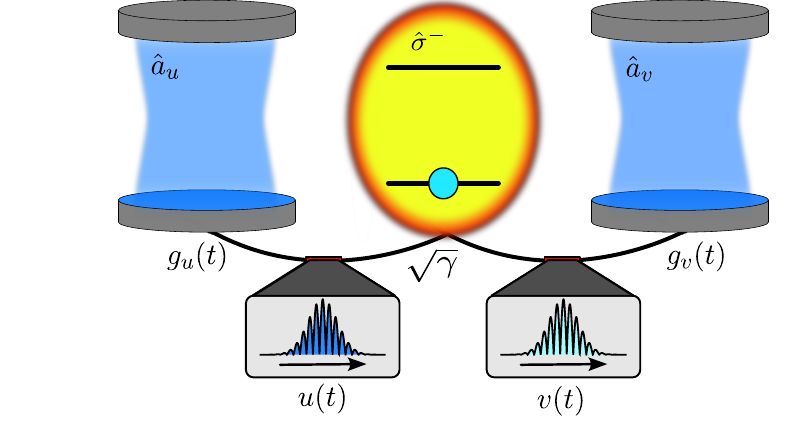}
\caption{Single-pulse cascade of the SUPER excitation scheme.}\label{fig.schem_2}
\end{figure}
We choose the selected output mode to be identical to the input mode, $v(t)=u(t)$. The corresponding single-pulse cascade is shown in Fig.~\ref{fig.schem_2}, and the SLH network is
\begin{align}
    G^{\text{s}} = G_v\lhd G_{\text{TLS}}\lhd G_u,\label{eq.G_cas2}
\end{align}
with cavity triplets $G_i = (\mathds{1},g_i(t)\hat{a}_i,0)$ for $i\in\{u,v\}$ and emitter triplet $G_{\text{TLS}}=(\mathds{1},\sqrt{\gamma}\hat{\sigma}^-,-\Delta\hat{\sigma}^+\hat{\sigma}^-)$. The resulting operators are
\begin{align}
    \hat{L}^{\text{s}} &= \sqrt{\gamma}\hat{\sigma}^- + \boldsymbol{g}^{\dagger}(t)\boldsymbol{a} \label{eq.Lind_cas2}
\end{align}
\begin{align}
    \hat{H}^{\text{s}} &= \hat{H}_{\text{TLS}} + \frac{i}{2}\boldsymbol{a}^{\dagger}\boldsymbol{A}(t)\boldsymbol{a}- \frac{\sqrt{\gamma}}{2i}(\hat{\sigma}^+\tilde{\boldsymbol{g}}^{\dagger}(t)\boldsymbol{a} - \text{h.c.}),\label{eq.Ham_cas2}
\end{align}
where $\boldsymbol{a} = (\hat{a}_{u},\hat{a}_{v})^T$, $\boldsymbol{g}=(g^*_{u},g^*_{v})^T$, $\tilde{\boldsymbol{g}}=(g^*_{u},-g^*_{v})^T$, and
\begin{align}
  \boldsymbol{A}(t) = \begin{pmatrix}
      0 & g_{u}^*(t)g_{v}(t) \\
      -g_{u}(t)g_{v}^*(t) & 0
  \end{pmatrix}.
\end{align}
The initial state is 
\begin{align}
    \ket{\psi_0}=\ket{\alpha}_{u}\otimes\ket{g}_{\text{TLS}}\otimes\ket{0}_{v},
\end{align}
and we solve the dynamics with the cumulant expansion. As in the main text, we characterize the nontrivial scattering contribution by the loss/gain function
\begin{align}
    \langle \Delta \hat{n}(t)\rangle = \langle \hat{n}_{u}(t)\rangle + \langle \hat{n}_{v}(t)\rangle - \langle \hat{n}_{u}(0)\rangle.
\end{align}
The results are shown in Fig.~\ref{fig.res_cas2}. In contrast to the two-pulse representation in the main text, this description does not directly reveal the underlying $(-2,+1)$ exchange between the two spectral components. It is therefore useful as a compact alternative representation of the drive, but it does not expose the multiphoton structure of the SUPER process.

\end{document}